\documentclass[letterpaper,10 pt,conference]{IEEEtran}

\usepackage{subfig}
\usepackage{amsmath}
\usepackage{amssymb}
\usepackage{amsfonts}
\usepackage{amsthm}
\usepackage[normalem]{ulem} 
\usepackage{mathtools} 
\usepackage{times}
\usepackage{algorithm} 
\usepackage{algpseudocode} 
\usepackage[usenames]{color}
\usepackage{verbatim}
\usepackage{array}
\usepackage{ellipsis}
\usepackage{url}
\usepackage{mathcomp} 
\usepackage{units} 

\usepackage[bookmarks=false]{hyperref} 
\usepackage[numbers,sort&compress]{natbib}
\usepackage{cleveref} 

\usepackage{tikz}
\usetikzlibrary{shapes,arrows}

\usepackage{bm}

\usepackage{tabularx} 

\DeclareMathOperator{\diag}{diag}

\newlength{\termone}
\newlength{\termtwo}
\theoremstyle{plain}
\theoremstyle{definition} 

\theoremstyle{remark}

\providecommand{\abs}[1]{\lvert#1\rvert}
\providecommand{\norm}[1]{\lVert#1\rVert}

\Crefformat{equation}{Eq.~(#2#1#3)} 
\Crefrangeformat{equation}{Eqs.~#3(#1)#4--#5(#2)#6} 

\begin{document}

\title{Joint Transmission with Dummy Symbols for Dynamic TDD in Ultra-Dense Deployments}

\author{Haris~Celik and~Ki~Won~Sung
\\
KTH Royal Institute of Technology, Wireless@KTH, Stockholm, Sweden\\
E-mail: \{harisc, sungkw\}@kth.se} 


\maketitle

\begin{abstract}
Dynamic time-division duplexing (TDD) is considered a promising solution to deal with fast-varying traffic often found in ultra-densely deployed networks. At the same time, it generates more interference which may degrade the performance of some user equipment (UE). When base station (BS) utilization is low, some BSs may not have an UE to serve. Rather than going into sleep mode, the idle BSs can help nearby UEs using joint transmission. To deal with BS-to-BS interference, we propose using joint transmission with dummy symbols where uplink BSs serving uplink UEs participate in the precoding. Since BSs are not aware of the uplink symbols beforehand, \emph{any} symbols with zero power can be transmitted instead to null the BS-to-BS interference. Numerical results show significant performance gains for uplink and downlink at low and medium utilization. By varying the number of participating uplink BSs in the precoding, we also show that it is possible to successfully trade performance in the two directions.
\end{abstract}


\section{Introduction}
\label{sec:intro}
Densification with small-cells is considered a promising solution in bringing about more capacity and increasing end user data rates \citep{6171992}. The premise of small cells is based on the notion that they are cheaper and easier to deploy thanks to lower transmit powers and therefore less restrictive regulations on the cell planning. This means that they can be placed where the actual traffic is and thereby increase quality of service. If they are equipped with idle mode capability, they can go to sleep mode and conserve energy.

For a given coverage area and fixed number of UEs, densification leads to smaller cell sizes and fewer UEs per BS. A limiting factor of densification is that the bandwidth reuse gain is exploitable only up to a certain point \citep{7037588}\cite{7343923}. As more and more BSs are deployed, bandwidth allocated to each UE increases linearly with the number of offloaded UEs. Eventually, as there are no more UEs to offload, the bandwidth reuse gain drops to zero and the network is said to be ultra-dense. Beyond this point, given everything else being equal, the capacity is governed by the spectral efficiency rather than spectrum reuse gain if one was to densify further.

Fast-changing traffic is expected to become more common in ultra-dense deployment due to the fact that the demand in each cell is driven only by a few UEs. To accommodate these fluctuations, adaptive time-resource allocation in the form of dynamic TDD has been proposed in literature. Dynamic TDD allows time slots to be allocated on a very short time scale for downlink and uplink transmissions, but is also prone to harsher interference conditions as so-called same-entity interference (UE-to-UE and BS-to-BS) is generated, in addition to existing other-entity-interference (BS-to-UE and UE-to-BS).


One implication of the ultra-densification and resulting low UE-to-BS density is that some BSs will not have an active UE to serve. Rather than going into sleep mode the idle BSs can be used to improve the performance of active UEs in nearby cells through joint transmission (JT), which is the underlying premise of this work. This implies that the same data can be transmitted by multiple BSs as long as they are part of the same cooperative set. JT is therefore likely to increase overall BS resource utilization and improve the signal-to-interference and noise ratio (SINR) for some UEs, while others not included in the cooperation may instead experience more interference and a decrease in their SINR. How to best utilize the JT in systems with same- and other-entity interference is an outstanding problem and the primary focus of this work.


Linear precoder design based on minimizing the mean-square error for a multi-cell multi-user multiple-input multiple-output (MIMO) system operating in dynamic TDD mode was treated in~\citep{7247176}\citep{7227096}. To the authors' knowledge, they appear to be one of the few works considering JT (and joint reception) with dynamic TDD. The multiple transmit and receive antennas in each cell enabled precoders and decoders to be constructed in a distributed per-cell fashion for both uplink and downlink, but may yield limited gains when BSs and UEs are equipped with a single antenna. It did also not address the aspect of inter-cell interference. Furthermore, because the system is distributed, it may not be able to fully utilize the spatial BS diversity when UE diversity is low and some BSs do not have a UE to serve.

In this paper, we consider network-wide JT where BSs cooperate to create a geometrically distributed antenna array in the downlink. Single antennas are assumed for BSs and UEs alike. This assumption stems from the fact that multi-antenna devices require multiple radio frequency chains which for various reasons (cost, form factor) can be difficult to implement in today's small cells. JT is facilitated using zero forcing which effectively nulls BS-to-UE interference and is limited by the number of antennas. Still, same-entity interference characteristic of dynamic TDD remains an issue. This aspect has been a point of interest in many previous works as same-entity interference can generate challenging interference to UEs and severely degrade their performance. To deal with this, literature proposes clustering where the same switching point is applied to all cells belonging to the same cluster. The switching point is set based on the traffic demand in each cluster rather than the whole network. While this eliminates same-entity interference inside each cluster, inter-cluster interference is still a problem. Other works considered more traditional interference management techniques such as power control or/and scheduling, or receiver-based interference cancellation/rejection. 

\begin{figure}[t!]
\centering
\includegraphics[width=.30\textwidth]{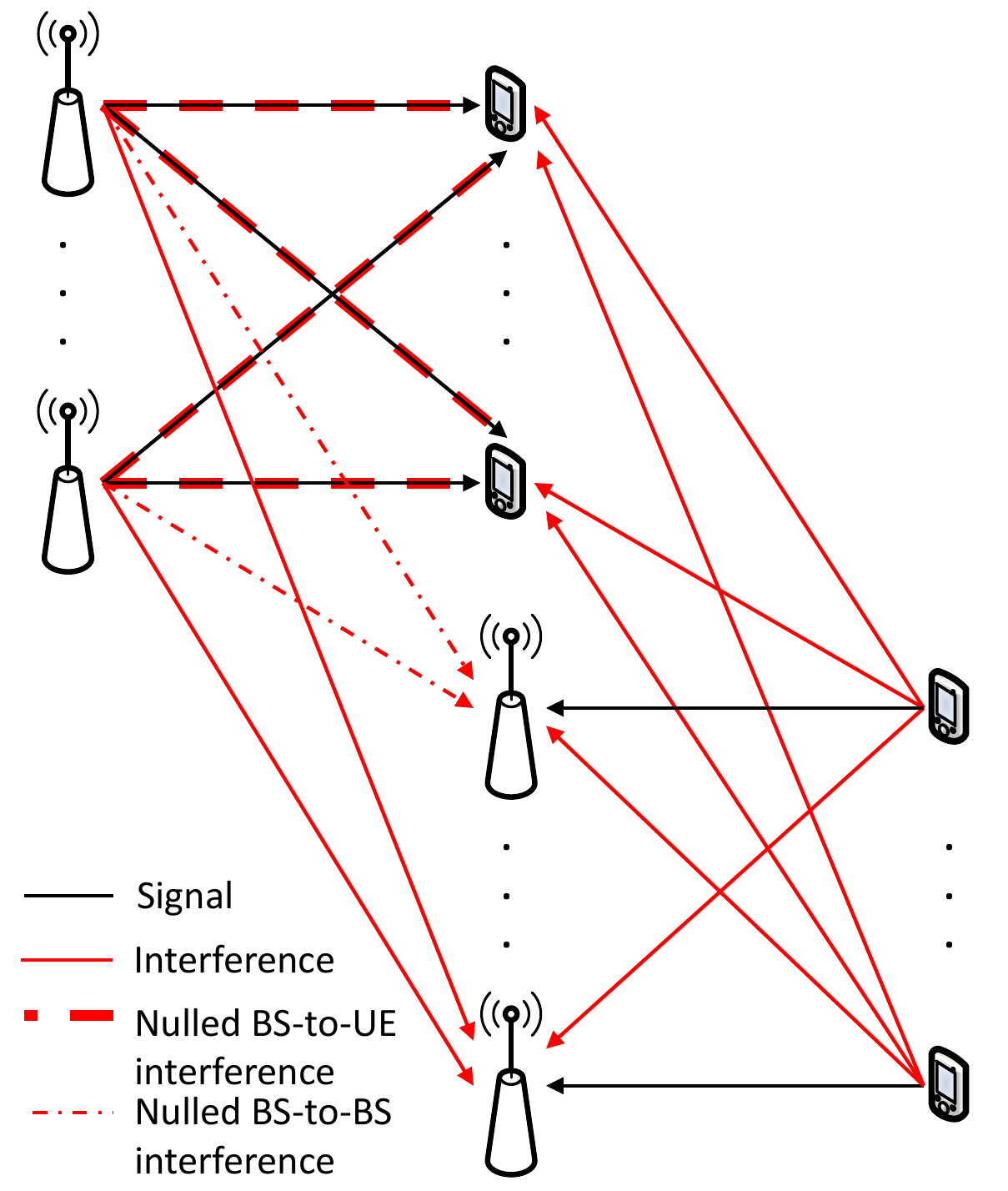}
\caption{Multi-cell dynamic TDD network with JT-DS.}
\label{fig:jtdsConcept}
\end{figure}


This paper includes uplink BSs in the precoder design to mitigate some of the BS-to-BS interference. Precoder coefficients may therefore be different compared to JT which only considers downlink UEs. BS-to-BS interference is considered one of the main limiting aspects of dynamic TDD in traditional deployments as the much larger downlink power and chances of line-of-sight tends to saturate uplink performance. The number of uplink BSs that can participate depends on the amount and mix of uplink and downlink traffic, and is constrained by the number of cooperating downlink antennas. Since downlink BSs are not aware of the information to be transmitted by uplink UEs beforehand, \emph{any} symbols with zero power can be transmitted to uplink BSs virtually. We denote this scheme as joint transmission with dummy symbols (JT-DS). In comparison, \citep{7070656} employs an asymptotically optimal precoder based on sum-power minimization in the context of massive MIMO for nulling downlink macro-tier interference to small-cell uplink BSs. Furthermore, we consider downlink-only power control which can be formulated as a convex optimization problem. In contrast, uplink UEs transmit with maximum power independently. This way, most of the complexity and coordination for the interference mitigation is introduced on the BS side. 


\section{System Model}
\label{sec:model}
For notational ease and without loss of generality, let $\mathcal{K}_{\text{dl}}=\{1,\ldots,K_{\text{dl}}\}$ and $\mathcal{K}_{\text{ul}}=\{K_{\text{dl}}+1,\ldots,K\}$ be the set of scheduled UEs where the subscript indicates their transmission direction. Similarly, let $\mathcal{N}=\{1,\ldots,N\}$ be the index set of all BSs in the network, including potentially idle ones. The disjoint subsets $\mathcal{N}_{\text{dl}}$ and $\mathcal{N}_{\text{ul}}$ denote BSs devoted to downlink transmission and uplink reception, respectively, where  $\abs{\mathcal{N}_{\text{dl}}}+\abs{\mathcal{N}_{\text{ul}}}=N_{\text{dl}}+N_{\text{ul}} \leq N$. 

The received signal in the downlink $(i\in\mathcal{K}_{\text{dl}})$ and uplink $(j\in\mathcal{K}_{\text{ul}})$ can be written as
\begin{align*}
y_i &= \mathbf{h}_{i}^H\mathbf{x} + \sum_{k=K_{\text{dl}}+1}^{K} \sqrt{P_u}g_{ik}s_k + n_i, \\
y_j &= \sum_{k=K_{\text{dl}}+1}^{K} \sqrt{P_u}h_{kb(j)}s_k + \mathbf{f}_{b(j)}^H\mathbf{x} + n_j,
\end{align*}
where $h_{ib(j)}$ denotes the channel between UE $i$ and BS $b(j)$ serving UE $j$, and $g_{ij}$ is the same-entity interference channel between downlink UE $i$ and uplink UE $j$. Defined is also channel vector $\mathbf{h}_{i}$ between downlink BSs and downlink UE $i$, and $\mathbf{f}_{b(j)}$ between downlink BSs and uplink BS $b(j)$. To simplify things, we assume perfect channel state information throughout this paper. Noise and data symbols are modeled complex Gaussian as $n_k\sim\mathcal{CN}(0,\sigma^2)$ and $s_k\sim\mathcal{CN}(0,1)$, respectively, and $P_u$ is the maximum UE transmit power.

\section{JT and JT-DS Concept}
The number of downlink UEs that can be included in the precoding is limited by the zero forcing constraint $K_{\text{dl}}\leq N_{\text{dl}}$. This, in turn, implies that the maximum number of uplink BSs that can be participate in the precoding is equal to
\begin{equation*}
V_{\text{ul}}^{\text{max}} = \min\left\{ N_{\text{ul}},N_{\text{dl}}-K_{\text{dl}} \right\}.
\end{equation*}
As we shall see, this number will inflict a trade-off between downlink and uplink performance as it affects the precoder design. Hence, the number of participating uplink BSs can be thought of as a system design parameter defined as:
\begin{equation*}
V_{\text{ul}}(\delta)=\max \left\{0, V_{\text{ul}}^{\text{max}}-\delta \right\}.
\end{equation*}
Exactly which uplink BSs that should participate is expanded upon in the next section.

Downlink and dummy symbols are mapped as 
\begin{equation*} 
\mathbf{x} = \mathbf{W}\mathbf{P}_d^{1/2}\mathbf{s}_{d},
\end{equation*}
where $\mathbf{s}_{d}=[s_1,\ldots,s_{K_{\text{dl}}},s_1^d,\ldots,s_{V_{\text{ul}}}^d]^T$ with dummy symbols $s_k^d$, $\mathbf{P}_{d}^{1/2}=\diag\left( \sqrt{p_1},\ldots,\sqrt{p_{K_{\text{dl}}}},\sqrt{p_{K_{\text{dl}}+1}},\ldots,\sqrt{p_{K_{\text{dl}}+V_{\text{ul}}}} \right)$, and $p_k=0, \forall k\in\mathcal{V}_\text{ul}=\{K_\text{dl}+1,\ldots,K_\text{dl}+V_\text{ul}\}$ representing zero power for dummy symbols. 


The transmit zero forcing precoding matrix is given as
\begin{equation} \label{eq:precodermatrix}
\mathbf{W} = \mathbf{M}^H(\mathbf{M}\mathbf{M}^H)^{-1},
\end{equation}
where $\mathbf{W}=[\mathbf{w}_1,\ldots,\mathbf{w}_{K_{\text{dl}}+V_{\text{ul}}}]$ with $\norm{\mathbf{w}_k}_2^2=1$ for scaling purposes in order to obtain feasible solutions with MATLAB Optimization Toolbox, and
\begin{equation*}
\mathbf{M}^H=
\begin{cases}
[\mathbf{h}_1^H,\ldots,\mathbf{h}_{K_{\text{dl}}}^H], & V_{\text{ul}}=0 \\
[\mathbf{h}_1^H,\ldots,\mathbf{h}_{K_{\text{dl}}}^H,\mathbf{f}_{b(K_{\text{dl}}+1)}^H,\ldots,\mathbf{f}_{b(K_{\text{dl}}+V_{\text{ul}})}^H], & V_{\text{ul}}\geq 1
\end{cases}
\end{equation*}
In general, ~\Cref{eq:precodermatrix} only holds when $K_\text{dl}+V_\text{ul}\leq N_\text{dl}$ and can be seen as the modified zero forcing constraint. 


The power constraint is expressed as
\begin{equation*}
\mathbb{E}\left[\abs{x_{n}}^2\right]=\sum_{k=1}^{K_\text{dl}+V_\text{ul}} \abs{w_{nk}}^2 p_k \leq P_b, \forall n\in\mathcal{N}_{\text{dl}}
\end{equation*}
From~\citep{4599181} we gather that the precoder in the form of the pseudo-inverse may be suboptimal for a per-antenna power constraint.

SINR for the received signal is
\begin{align*}
\gamma_i &= \frac{\abs{\mathbf{h}_{i}^H\mathbf{w}_{i}}^2 p_i}{\sigma^2 + \sum_{k=1,k\neq i}^{K_{\text{dl}}+V_{\text{ul}}} \abs{\mathbf{h}_{i}^H\mathbf{w}_{k}}^2 p_k +  \sum_{l=K_\text{dl}+1}^{K} \abs{g_{il}}^2 P_u}, \\
\gamma_j &= \frac{\abs{h_{jb(j)}}^2 P_u}{\sigma^2 + \sum_{l=K_\text{dl}+1,l\neq j}^{K} \abs{h_{lb(j)}}^2 P_u + \sum_{k=1}^{K_\text{dl}+V_{\text{ul}}} \abs{\mathbf{f}_{b(j)}^H\mathbf{w}_k}^2 p_k}. 
\end{align*}


Before proceeding, we emphasize some key aspects important to the applicability and effectiveness of JT-DS. Firstly, since UEs are equipped with a single antenna and not in a position to cooperate and share their information with each other, the ability to transmit jointly is confined to downlink BSs (the aspect of BS decoding is outside the scope of this work). Precoding can therefore be performed only when there is downlink traffic in the network $(K_{\text{dl}}\geq 1)$. All BSs not serving uplink traffic will take part in the joint transmission. In the absence of downlink traffic and same-entity interference, the network will operate distributedly in uplink mode. While downlink UEs can receive data from multiple downlink BSs, uplink UEs transmit only to their strongest BS without regard to other ongoing communication. Secondly, to mitigate BS-to-BS interference, JT-DS requires a minimum of uplink traffic $(K_{\text{ul}}\geq 1)$. If only downlink traffic is present, JT-DS becomes identical to JT. This is also the case if traffic is too high $(K=N)$ as it implies that $V_{\text{ul}}=0$. Thus, JT-DS requires a mix of both downlink and uplink traffic in less than fully loaded networks to be even considered, as illustrated in Figure \ref{fig:jtdsConcept}. Put together, these constraints may limit the applicability of the proposed scheme when multi-user diversity is very low, as is often the case in ultra-dense networks.

Assuming the above-mentioned conditions are satisfied $(V_{\text{ul}}\geq 1)$, it remains to determine which uplink BSs to include in the precoding in order to maximize sum-rate
\begin{equation*}
\sum_{k=1}^{K} B\log(1+\gamma_k)
\end{equation*}
where $B$ denotes bandwidth. Intuitively, by mitigating interference to worst performing UEs we can expect uplink performance to improve. Again, without loss of generality, let $\gamma_{K_\text{dl}+1}^b\leq\cdots\leq \gamma_{K}^b$ be the uplink SINRs for the baseline scheme~\citep{7145967} where not only uplink UEs but also downlink BSs transmit independently. To increase sum-rate, BSs with index $b(K_\text{dl}+1),\ldots,b(K_\text{dl}+V_\text{ul})$ corresponding to the $V_{\text{ul}}$ lowest uplink rates are included in the precoder design. This selection process is suboptimal as picking poorly performing UEs subject to a bad propagation environment rather than interference should give no uplink gain with JT-DS versus JT.


\subsection{Downlink power control}
We also consider power control in the downlink for JT and JT-DS as much more complexity can be handled by the BS. Transmit powers are based on maximizing downlink sum-rate and obtained by solving
\begin{equation}
\newlength\subto
\settowidth{\subto}{$\scriptstyle p_1,\ldots,p_{K_\text{dl}+V_\text{ul}}$}
\begin{aligned}
& \underset{p_1,\ldots,p_{K_\text{dl}+V_\text{ul}}}{\text{maximize}}
& & \sum_{k=1}^{K_\text{dl}+V_\text{ul}} \log(1+\gamma_k) \\
& \text{\makebox[\subto]{\text{subject to}}}
& & \sum_{k=1}^{K_\text{dl}+V_\text{ul}} \abs{w_{nk}}^2 p_k \leq P_b, \, \forall n\in\mathcal{N}_\text{dl} \\
& & & p_k\geq 0, \, \forall k\in\mathcal{K}_\text{dl} \\
& & & p_k= 0, \, \forall k\in\mathcal{V}_\text{ul}
\end{aligned}
\tag{P1}
\label{opt:P1}
\end{equation}
where $P_b$ is the maximum BS transmit power. Because uplink transmissions take place independently with no coordination, we relax the objective function in \eqref{opt:P1} so that it excludes UE-to-BS interference. Moreover, since transmit powers for the dummy symbols are zero, optimizing over only the first $K_{\text{dl}}$ variables does not affect the total BS power budget. The zero powers imply that $R_k=0$ for all $ k\in\mathcal{V}_\text{ul}$. Assuming the BS-to-UE interference is perfectly nulled, \eqref{opt:P1} equivalates to

\begin{equation}
\begin{aligned}
& \underset{p_1,\ldots,p_{K_\text{dl}}}{\text{maximize}}
& & \sum_{k=1}^{K_\text{dl}} \log_2(1+p_k) \\
& \text{subject to}
& & \sum_{k=1}^{K_\text{dl}} \abs{w_{nk}}^2 p_k \leq P_b, \, \forall n\in\mathcal{N}_\text{dl} \\
& & & p_k\geq 0, \, \forall k\in\mathcal{K}_\text{dl}
\end{aligned}
\tag{P2}
\label{opt:P2}
\end{equation}
This problem is convex since the objective function is concave and the constraints are linear in $p_1\ldots,p_{K_\text{dl}}$. To improve runtime, the objective function in \eqref{opt:P2} is replaced with $\sum_{k=1}^{K_\text{dl}} p_k$ in our simulations so that it can be solved as a linear program. It is noted that \eqref{opt:P2} is identical to the downlink power control problem for JT, though precoders $\{w_{ij}\}$, and consequently SINRs, are in general different.


\begin{table}[t]
\caption{Simulation parameters.}
\centering
\begin{tabular}{ p{3.9cm} | p{3.9cm} }
\hline
Number of BSs & 16 \\ 
Number of UEs per BS & $\leq 1$ \\ 
Channel model & Average path loss of WINNER II A1 \citep{WINNER07:D1.1.2} with fast fading \\ 
Bandwidth & 10 MHz \\ 
Frequency & 2 GHz \\ 
UL:DL traffic demand & 50:50 \\ 
BS and UE transmit power & $\leq$ 100 mW \\ 
Frequency reuse & Universal \\ 
Noise figure & 9 dB \\ 
Traffic model & Infinitely backlogged \\ 
UE distribution &  Uniform \\ 
Total number of snapshots & 10 000 \\ \hline
\end{tabular}
\label{tab:params}
\end{table}

\section{Numerical results}

Monte Carlo simulations are performed for the statistics collection based on the parameters in Table~\ref{tab:params}. BSs are placed according to a grid deployment in an indoor open area spanning 40x40 meters. Local scattering in terms of fast (Rayleigh) fading is also included in the propagation model. This scenario is interesting since indoor short-range communication allows for similar transmit powers to be used which has shown promising results for dynamic TDD at low and medium traffic load \citep{Jänis2012}. Long-term traffic demand is evenly distributed between uplink and downlink, though instantaneous traffic can be highly asymmetric. As a baseline we consider~\citep{7145967} where not only uplink UEs but also downlink BSs transmit independently. The lack of coordination of \citep{7145967} means it provides near lower bound performance, yet it has shown to perform well in an ultra-dense setting compared to TDD systems with a static switching point. System utilization signifies the relative load in the network $(K/N)$. For worst user performance we consider the 5th percentile sum-rate of all realizations divided by the traffic load $K$.




\begin{figure}[t!]
\centering
\includegraphics[width=.50\textwidth]{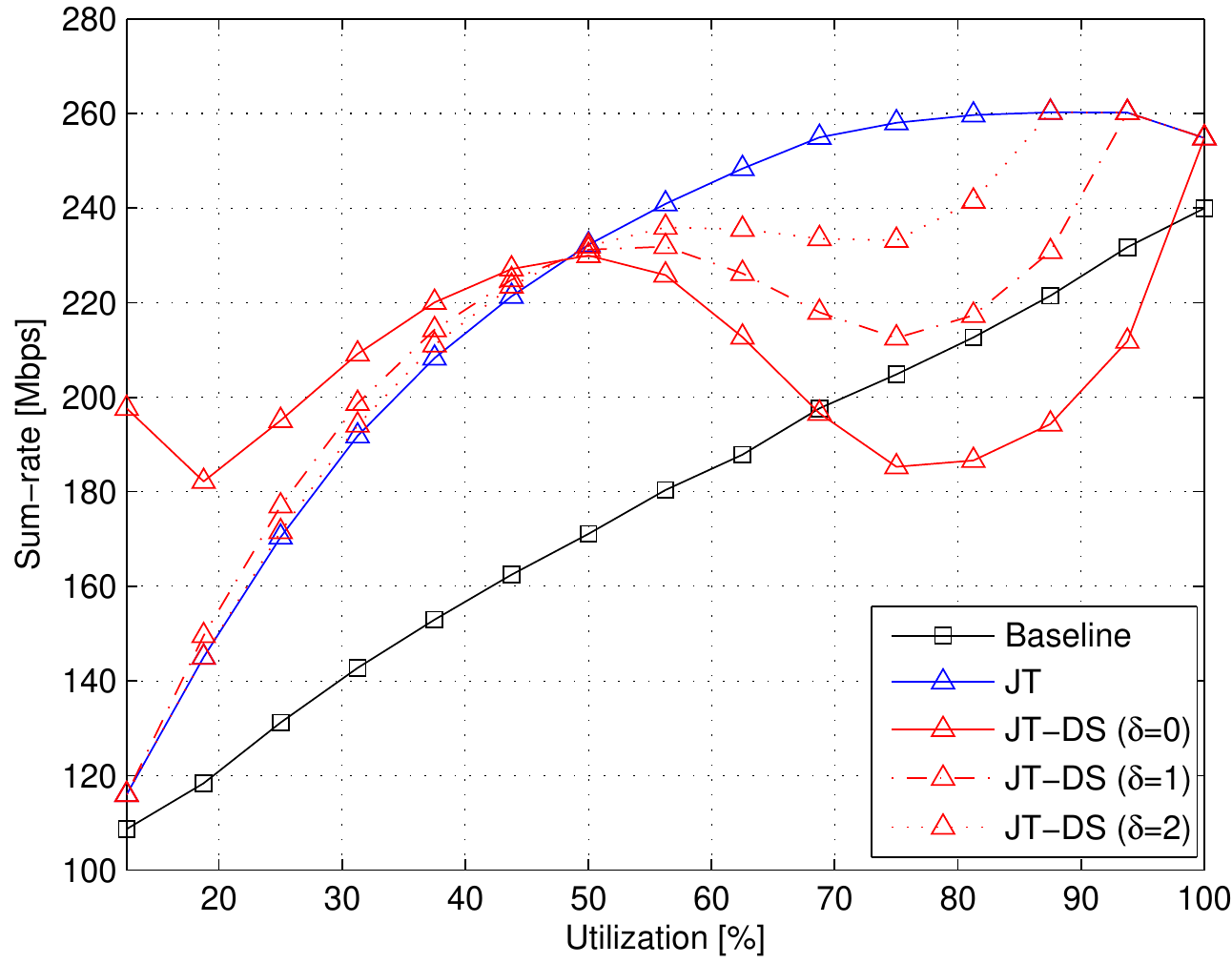}
\caption{Average sum-rate in the presence of both uplink and downlink traffic.}
\label{fig:avgSumRate}
\end{figure}

\begin{figure}[t!]
\centering
\includegraphics[width=.50\textwidth]{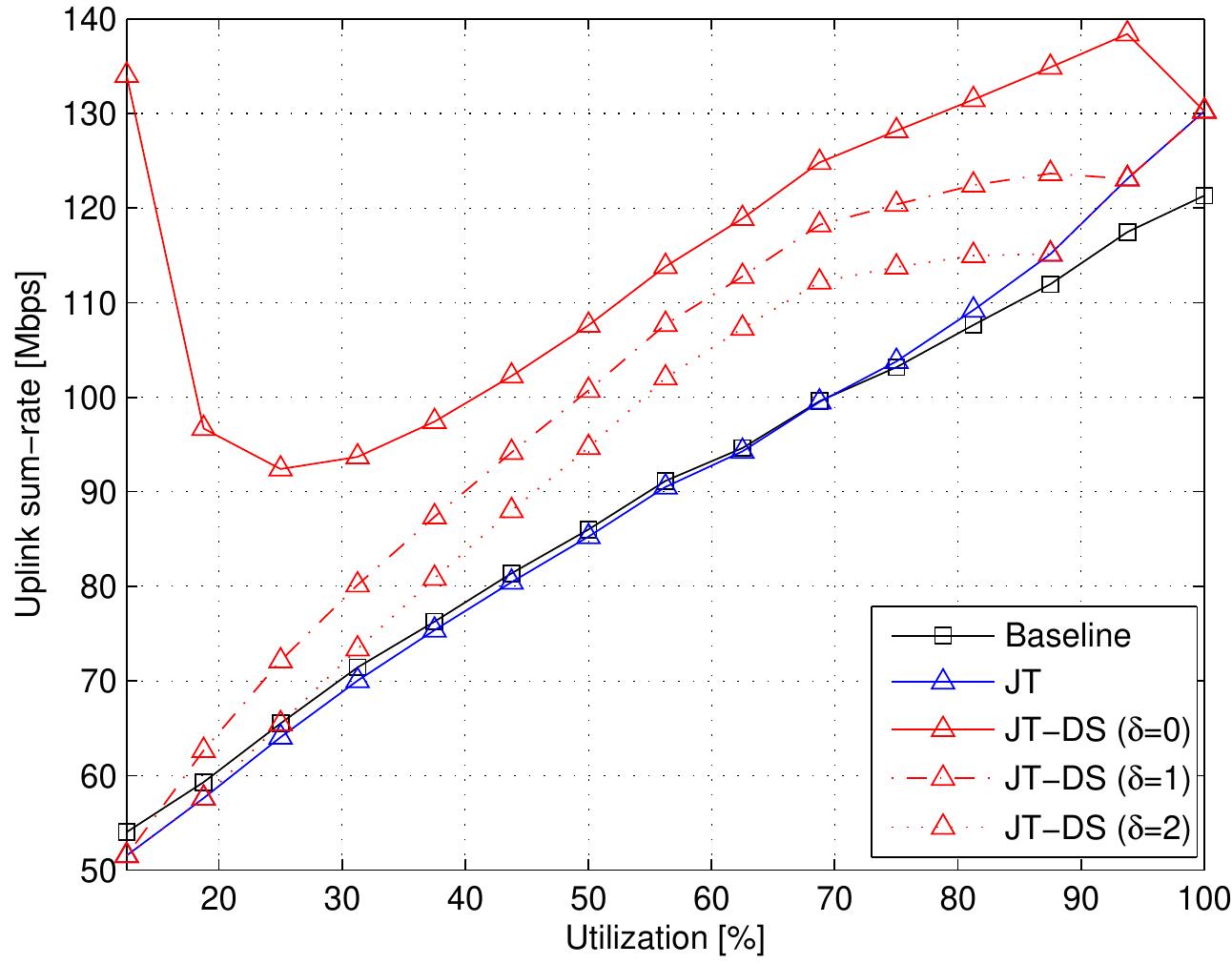}
\caption{Average uplink sum-rate in the presence of both uplink and downlink traffic.}
\label{fig:avgUlSumRate}
\end{figure}

\begin{figure}[t!]
\centering
\includegraphics[width=.50\textwidth]{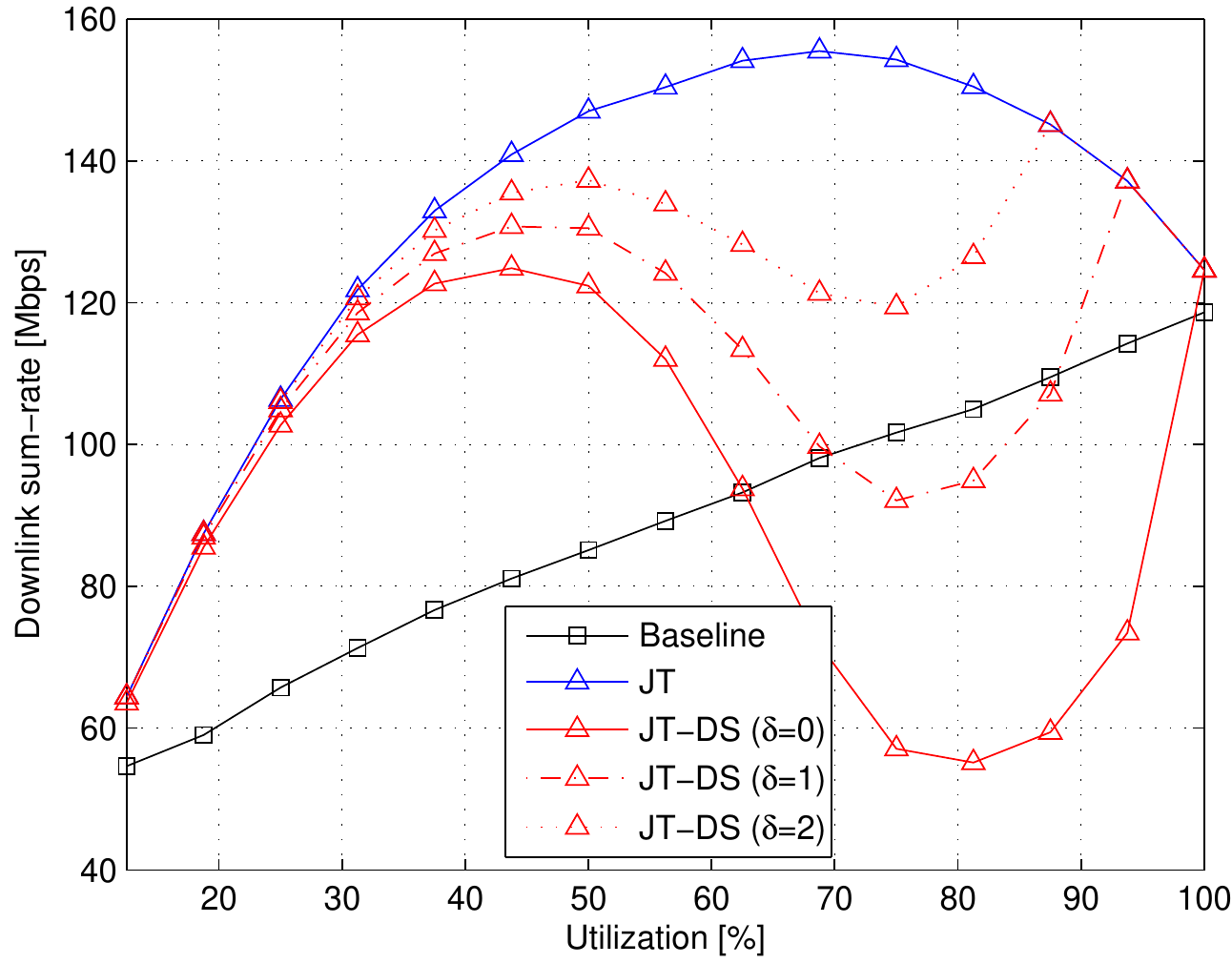}
\caption{Average downlink sum-rate in the presence of both uplink and downlink traffic.}
\label{fig:avgDlSumRate}
\end{figure}

\begin{figure}[t!]
\centering
\includegraphics[width=.50\textwidth]{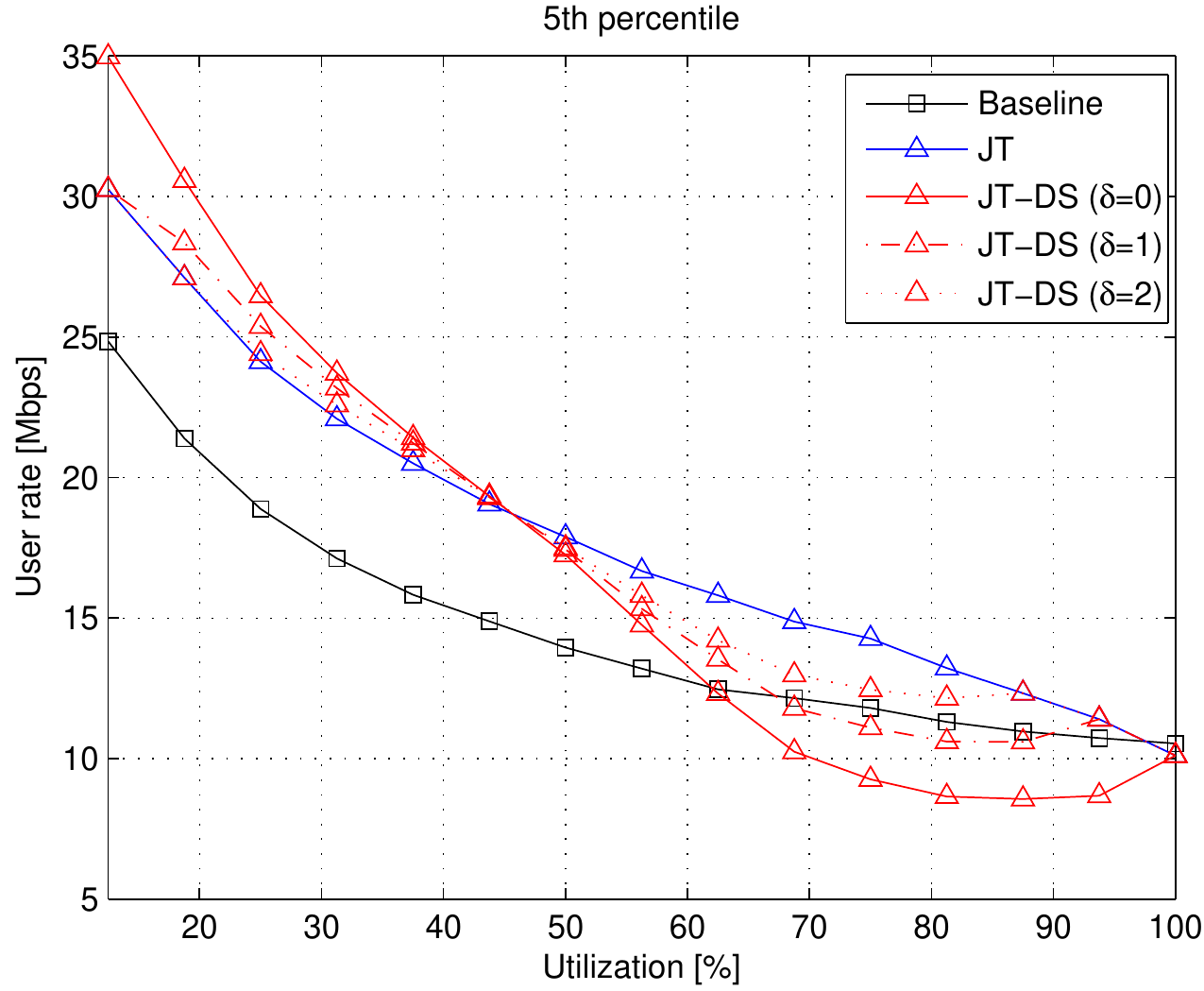}
\caption{5th percentile user rate in the presence of both uplink and downlink traffic.}
\label{fig:5thpercSumRate}
\end{figure}

Average sum-rate for JT and JT-DS is illustrated in Figure \ref{fig:avgSumRate} together with the baseline scheme. Corresponding uplink and downlink sum-rate is shown in Figure \ref{fig:avgUlSumRate} and \ref{fig:avgDlSumRate}, respectively. In order to adequately evaluate JT-DS, we only consider the case when there is both uplink and downlink traffic in the network. The lowest utilization point which reflects an interference-free environment and a single user is therefore omitted. Because the interference is lower at low utilization, the sum-rate may drop initially until traffic demand in the network is high enough to counter the decrease. At low traffic load and utilization, it is shown that JT and JT-DS provides a substantial performance gain in the downlink thanks to the nulling of BS-to-UE interference, though uncontrolled UE-to-UE interference limits further gains. In the uplink, JT-DS significantly improves sum-rate by including uplink BSs in the precoding which reduces BS-to-BS interference, but similar to the downlink case it is constrained by uncontrolled UE-to-BS interference. Worst (5th) percentile user performance is depicted in Figure \ref{fig:5thpercSumRate}. 



As more and more UEs are activated and utilization increases, so does unmitigated UE-to-BS and UE-to-UE interference. In the uplink, increasing utilization implies more uplink (and downlink) traffic and, in turn, initially increasing $V_{\text{ul}}$. More uplink BSs are thus able to participate in the precoding. Consequently, uplink sum-rate increases in part thanks to higher uplink demand, and in part thanks to the fact that more of the BS-to-BS interference is mitigated. At some utilization point however, $V_{\text{ul}}$ will instead begin do decrease, meaning fewer participating uplink BSs. Despite this, an uplink performance gain is still achievable by including uplink BSs corresponding to worst performing uplink UEs. In downlink, however, performance of JT-DS starts to diminish as adding more receivers also increases ill-conditioning of the precoder matrix. As a result, elements $\{\abs{w_{ij}}^2\}$ may become overly large, and transmit powers are lowered to compensate for the difference in order to not violate the BS power constraint. This is observed for both JT and JT-DS in Figure \ref{fig:avgDlSumRate}, though the difference is more accentuated for JT-DS as the participating uplink BSs do not contribute in increasing downlink sum-rate, in addition to the more severe ill-conditioning. Downlink performance can therefore become a bottleneck for JT-DS at higher traffic loads. On the other hand, the lower downlink powers help reduce BS-to-BS interference and improve uplink performance. At full utilization, JT-DS becomes identical to JT since $V_{\text{ul}}=0$. 


Because of the ill-conditioning, we expect a trade-off between downlink and uplink performance for JT-DS when varying the size of the precoder through $\delta$. When utilization is low, the effects of increasing $\delta$ are especially noticeable in the uplink where the number of uplink BSs is already small due to an exceedingly low UE diversity. This implies that $V_{\text{ul}}$ will be close or equal to zero even for small values of $\delta$. As a result, few if any uplink BSs will be able to participate in the precoding. In downlink, however, the size of the precoder is still fairly small and the effects of ill-conditioning less severe compared to high utilization regime. In contrast, at high utilization even small changes in the number of participating uplink BSs can make a large difference to downlink performance. At the same time, it is evident that including fewer uplink BSs in the JT-DS will effectively increase BS-to-BS interference to uplink BSs excluded from the precoding and lower overall uplink sum-rate.

\section{Conclusion}
This paper considered network-wide JT for dynamic TDD in ultra-dense small-cell deployment where single antenna BSs cooperate in the downlink. Zero forcing precoding is facilitated to effectively null BS-to-UE interferences. To also deal with BS-to-BS interference characteristic of dynamic TDD, JT with dummy symbols is introduced by including uplink BSs in the precoder design. 

The proposed scheme is shown to significantly improve both uplink and downlink performance at low and medium utilization. In high utilization regime, an uplink performance gain is still achievable thanks to the inclusion of fewer but more interference-prone uplink BSs in the precoding. At the same time, downlink performance diminishes from ill-conditioning as more downlink UEs are added, even as the number of participating uplink BSs decreases. It is also shown that by varying the number of participating uplink BSs in the precoding, it is possible to trade uplink and downlink performance with each other and thereby improve downlink performance also at high utilization.

\section*{Acknowledgement}
Part of this work has been performed in the framework of the H2020 project METIS-II co-funded by the EU. Authors would like to acknowledge the contributions of their colleagues from METIS-II although the views expressed are those of the authors and do not necessarily represent the views of the METIS-II project.

\bibliographystyle{IEEEtran}
{\footnotesize\bibliography{IEEEabrv,mybib}} 

\end{document}